\documentclass[11pt,a4paper]{article}

\usepackage[utf8]{inputenc}
\usepackage[T1]{fontenc}
\usepackage{amsmath}
\usepackage{amssymb}
\usepackage[margin=2.5cm]{geometry}
\usepackage[round]{natbib}
\usepackage{tabularx}
\usepackage{graphicx}
\usepackage{url}
\usepackage{subcaption}
\usepackage{algorithm}
\usepackage{algpseudocode}
\algblockdefx[Switch]{Switch}{EndSwitch}[1]{\textbf{switch} #1}{\textbf{end switch}}
\algblockdefx[Case]{Case}{EndCase}[1]{\textbf{case} #1}{\textbf{end case}}

\title{Who Decides? Agency and Legitimacy in Digital Educational Systems}
 \author{
   Eriam Schaffter\\ ERIC (UR 3083), Université Claude Bernard Lyon 1\\ \texttt{eriam.schaffter@etu.univ-lyon1.fr}
   \and
   Ahmed Bounekkar\\ ERIC (UR 3083), Université Claude Bernard Lyon 1\\ \texttt{ahmed.bounekkar@univ-lyon1.fr}
 }
\date{}

\begin{document}
\maketitle

\newcommand\blfootnote[1]{%
  \begingroup
    \renewcommand\thefootnote{}\footnote{#1}%
    \addtocounter{footnote}{-1}%
  \endgroup
}

\begin{abstract}

Digital Educational Systems (DES) enable millions of individuals to learn or to acquire new skills. Their promises vary from one platform to another: some grant genuine degrees, others pledge rapid progress, drawing for this purpose on a range of techniques, from the recommendation of learning resources through complex models to a plain course textbook uploaded onto a learning management system. In this article, we argue that categorizing DES along two dimensions allows to map differences in the way these DES structurally conceptualise the learning process. These underlying dimensions are at once dimensions of the student model and mechanisms built in the DES's designs; we identify them as agency and legitimacy. Indeed, DES are, on the one hand, anchored in an institutionalised regime that grants them legitimacy and, in turn, legitimises the learning undertaken on these platforms; on the other hand, they make design choices as to the share of decision left to the learner in the learning process. We map a curated sample of common DES onto a plane composed of these two dimensions translated into indicators. This allows us to highlight underpopulated regions of the plane and to envisage how a DES might regulate these two dimensions algorithmically, in the interest of learning.

\end{abstract}


\section{Introduction}
In this article we use the term Digital Educational Systems (DES) to designate a computing system whose object is learning and for which the computational medium is not an addendum. DES include Intelligent Tutoring Systems \citep{vanlehn_relative_2011, graesser_autotutor_2004}, Learning Management Systems, Educational Recommender Systems \citep{da_silva_systematic_2023, ricci_recommender_2022}, and any other digital system whose intent is to teach and whose medium is information technology. With the rise of digital technology, these systems have undergone considerable development \citep{ciarli_digital_2021}, and the academic community reflects, through the sheer number of conferences and journals directly concerned with these tools, just as many ways of conceiving, designing, or observing how they operate. This article sets out to defend the claim that two structuring, orthogonal dimensions run through the various approaches to DES design, and that examining existing approaches through the lens of these two dimensions reveals unexplored regions. While a DES can be analyzed through numerous technical or pedagogical lenses (e.g., user interface, social learning features, or specific AI algorithms), we argue that Agency and Legitimacy constitute the foundational, orthogonal structural core of any educational system and that they are implicitly or explicitly reflected both on the student model and on the tutor model \citep{vanlehn_behavior_2006}. They represent the fundamental socio-technical trade-off of institutionalized learning: the tension between the learner's need for autonomy \citep{deci_what_2000} and the institution's mandate to standardize and certify knowledge \citep{weber_economy_1978}. This two-dimensional lens is sufficient to map the macro-behavior of these systems, as it captures who holds the power of decision-making, which ultimately dictates the system's underlying algorithmic architecture.

Thus, at the very heart of designing a DES, we make decisions about agency and legitimacy on the learner's behalf. What agency, then, and what legitimacy are decided for the learner in the principal DES in use today, and by what means might these dimensions be regulated in a DES that would place them at its center ?

\section{A landscape of digital educational systems}

\subsection{Agency and legitimacy}

Agency, in psychology, is the capacity to act as an agent, that is, to initiate and control one's actions and to experience oneself as responsible for them \citep{bandura_social_2001, deci_what_2000}. Legitimacy, in political science, is the right of an authority to direct the actions of a society; and authority itself, in sociology, is socially acquired power \citep{weber_economy_1978}. 

\subsection{Characterizing a curated set of DES}

Thus, in Digital Educational Systems (DES) we can observe, through a set of characteristics, a system's propensity to offer the learner agency and to confer legitimacy. For this analysis we selected a set (n=30) of DES representative of tools drawn from both research and industry. The complete catalogue of the systems, the coding grid, and the public documentation source used to code each of them are provided as supplementary material, together with the per-system rationale for every coding decision and the notebook that regenerates the figures reported here \citep{schaffter_agency_2026}.

In order to map a set of DES, we define a collection of criteria that yield a representation of the catalogued systems. Specifically, we seek to compute, in a simple manner, one indicator of the agency and one indicator of the legitimacy afforded by a given DES. To this end, we define a set of sufficiently observable features from which the two indicators are derived. We retain the following features. For agency, the first feature characterises who selects the next activity \citep{deci_what_2000, deschenes_recommender_2020}; the second records the cardinality of the activity choice offered at each step \citep{deci_what_2000}; the third captures whether the goal is set by the learner or by the system \citep{zimmerman_becoming_2002}; and the fourth indicates whether the learner may interrupt an activity \citep{wrosch_adaptive_2003}. For the legitimacy axis, we first observe the assessment frequency \citep{black_assessment_1998}, then whether the system is aligned with official competency standards \citep{european_centre_for_the_development_of_vocational_training_cedefop_2025}; we next record the type of credential issued \citep{bills_credentials_2003} and, finally, the organisation that confers the accreditation \citep{weber_economy_1978}.

Then, each feature is rescaled so its lowest level is 0 and its highest is 1, and each axis score is the unweighted mean of its four rescaled features, yielding two indicators in [0,1]. So for a system $s$, each coded feature $f$ with $m_f$ ordered levels is mapped to $v_f(s)=j_f(s)/(m_f-1)\in[0,1]$, where $j_f(s)$ is the rank of its level. The levels, ordered from 0 to 1 and named as in the deposited data, are: A1 selects next (system, mixed, learner); A2 choice set (one, a few, free); A3 sets the goal (system, negotiated, learner); A4 may interrupt (no, yes); L1 cadence (never, end, milestones, continuous); L2 official standard (no, yes); L3 credential (none, record, badge, certificate, degree); L4 accreditation (no one, publisher, university, State).

The two composite indicators are the equal-weight means over each axis:
\begin{equation}
\mathrm{Agency}(s) = \frac{1}{4}\sum_{f\in\{A1,\dots,A4\}}v_f(s),
\qquad
\mathrm{Legitimacy}(s) = \frac{1}{4}\sum_{f\in\{L1,\dots,L4\}}v_f(s).
\label{eq:composite}
\end{equation}

\begin{figure}[ht]
\centering
\includegraphics[height=13cm]{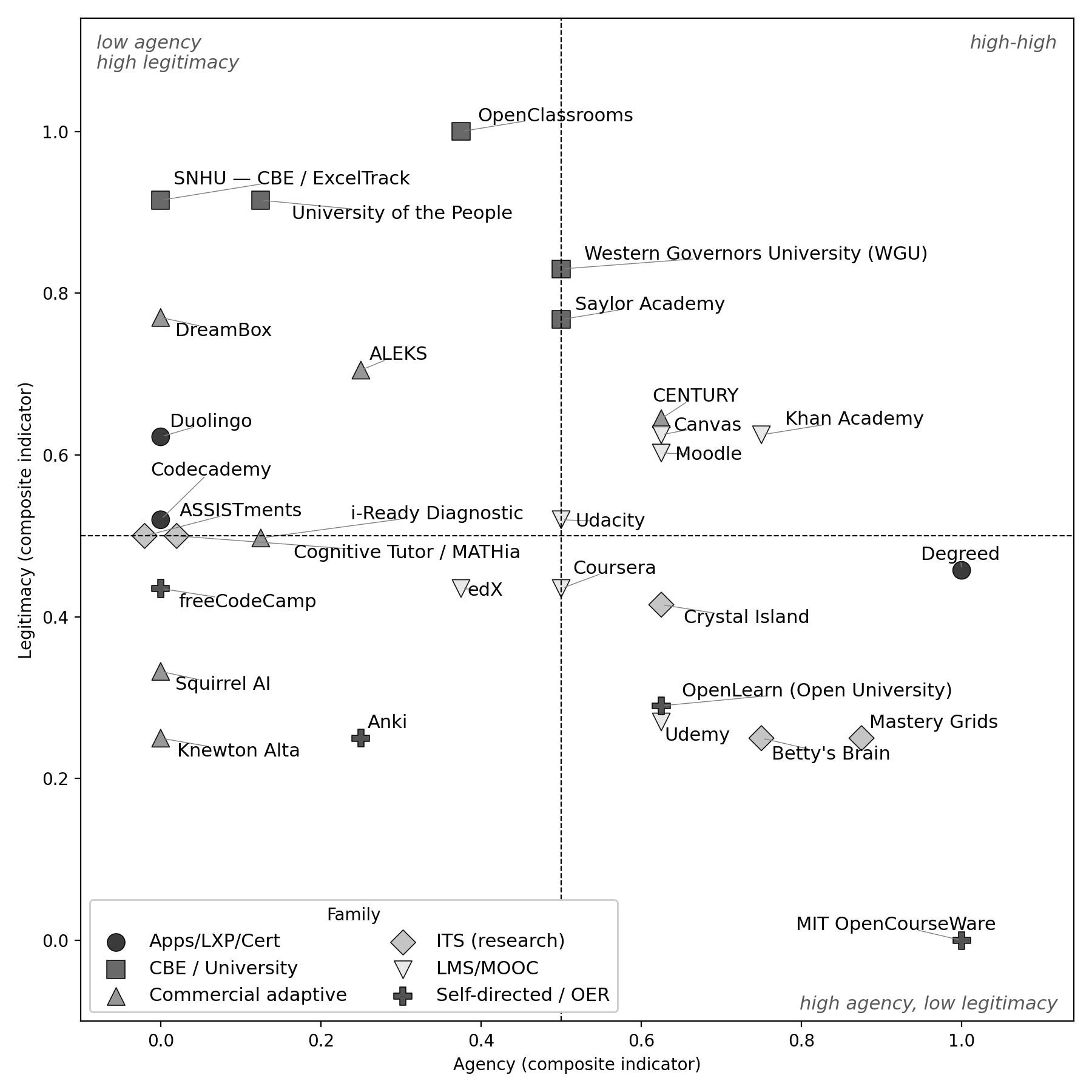}
\caption{Quadrant of digital educational systems showing empty region for high agency and high legitimacy.}
\label{fig:quadrant}
\end{figure}

So, when we calculate each system's coordinates with Equation \ref{eq:composite} we can represent the catalogued DES within a quadrant, shown in Figure \ref{fig:quadrant}, along the two axes of agency and legitimacy. The representation highlights a singular absence, in the upper-right corner, of systems offering both legitimacy and agency.

\subsection{Both from the learner and the system}

We take agency and legitimacy to be, at once, dimensions of the learner model \citep{vanlehn_behavior_2006, wang_development_2025} and of the tutor model. Regarding agency first, the learner is endowed with a dimension describing their capacity to be agentive in their learning and the DES with a capacity to regulate it. Likewise, regarding legitimacy which represents the degree of institutionalized confidence placed in their acquired competencies and know-how, a conceptualization close to assessment techniques like bayesian knowledge tracing \citep{corbett_knowledge_1995} or item response theory \citep{an_item_2014}, we also consider it a capacity of the DES, namely that of seeking to legitimise the student's knowledge through an approach that regulates assessment dispensation.

\subsection{Balancing agency and legitimacy}

What these representations also reveal is the possible absence, within these systems, of any continuous mechanism for balancing, over the course of a curriculum, the preservation of the learner's agency against the legitimacy that the curriculum ought to confer.

In order to balance agency and legitimacy, one must therefore envisage a DES that, through a model and in accordance with the criteria we used to map our DES, gives the learner choice, offers abundant resources, allows learners to set their own goals and to interrupt an activity, while simultaneously drawing on an official standard, assessing the learner frequently \citep{black_assessment_1998, roediger_test-enhanced_2006}, and awarding a high-quality credential issued by an official body. And while some of these criteria cannot be incorporated into a model because they are mostly exogenous (for instance, whether the DES is backed by an official institution), certain others can be. We can thus begin to glimpse what an algorithm for balancing agency and legitimacy within a DES might look like; we draft such a suggestion in Algorithm \ref{alg:balance}. 

This algorithm is compatible with any type of DES, provided that the DES offers the learner a choice among a set of educational resources, whether formative or evaluative. The resource consumed by the learner is thus chosen from a subset $R*$ of $R$, the set of available resources, whose composition and cardinality depend both on the observed agency and legitimacy and on the agency and legitimacy targeted for the learner. In the example algorithm let $S_a$ be the learner's agency and $S_L$ the learner's legitimacy, both kept in $[0,1]$. Each update is a bounded step of fixed size $d \in (0,1]$ passed through the saturation $\mathrm{sat}(x)=\min(1,\max(0,x))$, so the state never leaves $[0,1]$. The algorithm reads a window of recent traces and takes four calibration parameters: two thresholds above and below which the menu widens or tightens, a menu size, and a variety floor counted in distinct resource types.

\begin{algorithm}[t]
\caption{Updating agency and legitimacy from the consumed resource}
\label{alg:balance}
\begin{algorithmic}[1]
\State $k \gets m$ if $S_a \geq \tau_a$, else $k \gets 1$ \textit{ // } $S_a, S_L \in [0,1]$
\Repeat
    \State $R^\star \gets$ a menu of $k$ resources, evaluative ones first when $S_L < \tau_L$
    \State the learner consumes one resource of $R^\star$; $T_n$ is the resulting window
    \If{$T[-1]$ is evaluative and successful} $S_L \gets \mathrm{sat}(S_L + d)$
    \ElsIf{$T_n$ holds no evaluative resource} $S_L \gets \mathrm{sat}(S_L - d)$
    \EndIf
    \If{$T[-1]$ is evaluative and $T_n$ holds a formative resource}
    \State \hskip 1em $S_a \gets \mathrm{sat}(S_a + d)$ on success, $\mathrm{sat}(S_a - d)$ otherwise
    \EndIf
    \If{$T_n$ spans $v_{\min}$ distinct resource types or more} $S_a \gets \mathrm{sat}(S_a + d)$
    \Else{} $S_a \gets \mathrm{sat}(S_a - d)$
    \EndIf
    \If{mean rank of $T_n$ within $R^\star \geq k/2$} $S_a \gets \mathrm{sat}(S_a + d)$
    \EndIf
\Until{session ends}
\end{algorithmic}
\end{algorithm}

In the deliberately simple algorithm we propose, we reward demonstrated success on evaluative resources by increasing legitimacy, and let legitimacy erode when no evaluative resource has been consumed recently, reflecting the degradable nature of an institutional warrant. As for agency, we observe assessment outcomes so as to reward it when the learner follows successful trajectories, that is, by inspecting the traces, when we detect the consumption of formative resources followed by a successful evaluative one; conversely, we reduce agency when that same sequence ends in failure. We likewise reward agency when the window spans at least as many distinct resource types as the variety floor, and reduce it below that floor. Finally, we grant additional agency when the mean rank of the recently consumed resources sits in the lower half of the menu, that is, when the learner reaches past the items ranked first in the subset $R*$.

\section{Conclusion}

In this work, we positioned Digital Educational Systems (DES) to bring to light a gap these systems do not yet fill. Indeed, DES are distributed fairly evenly across the agency and legitimacy offered to the learner, the system's user, with the exception of the desirable but unpopulated region that would grant the learner both high agency and high legitimacy.

Surveying this landscape, we observe that the learner has several options and that these options engage the learner in fairly characteristic learning processes. First, providers of degree-granting programmes, issuing credentials certified by well-established institutions: this group offers programmes low in agency, whose counterpart is strong institutional legitimacy. Second, course marketplaces: they let the learner choose their own pace and learning activities, but offer little legitimacy and, above all, suffer from high attrition \citep{kizilcec_self-regulated_2017}. Third, open personal-exploration tools: they provide abundant learning resources and a diffuse legitimacy, sometimes through badges, resting essentially on their reputation \citep{bills_credentials_2003}. Finally, skill-building tools, highly directive in nature: they offer little agency but allow the learner to obtain, in a tightly paced manner, a degree of recognition for precise, well-defined competencies.

If the literature tells us that agency matters for learning \citep{deci_what_2000, zimmerman_becoming_2002, bandura_social_2001}, it appears essential that the DES supported by educational institutions, as represented on this quadrant, should adapt so as to provide learners with platforms offering both legitimacy and agency. This would thus account for the orthogonality of these two dimensions, an orthogonality that holds by construction as long as no adequate response is given to the dilemma between self-legitimation outside the institution and the difficulty of affording the learner a degree of freedom within an institutional framework.

To this end, we have proposed in this paper a conceptualisation of a DES incorporating an algorithm that would balance the agency afforded to the learner against the legitimacy conferred by the assessment of their competencies and knowledge. This positioning work resonates with a line of research on trajectory-based educational recommender systems formalised through control theory \citep{schaffter_trajectory-based_2025,memarian_education_2025,pasquale_recommender_2026}. In this formalisation, which places the learning trajectory at the heart of the model, educational practices apply fully to guide the learner in a constrained manner, while still leaving to the learner, at each learning step, the decision as to which educational resources to engage with. Within that programme, the balance between agency and legitimacy occupies a central place, and the model is what gives institutions and learners confidence in it.

\section*{Limitations}

Coding was done in a single pass by a large language model (Claude, Anthropic), applying the deposited rubric to each system's public documentation, and reviewed by the authors. Single-pass, single-coder classification is a limitation. The rubric, the sources and the rationale for all 240 decisions are deposited, so the coding can be audited.

\section*{Declaration of generative AI use}

During the preparation of this work, the authors used Claude (Anthropic) for curation and annotation, code generation, rephrasing and pre-submission review. The authors reviewed and edited all output and take full responsibility for the published content.

{\footnotesize
\bibliographystyle{apalike}
\bibliography{references}   

\begin{thebibliography}{}

\bibitem[An and Yung, 2014]{an_item_2014}
An, X. and Yung, Y.-F. (2014).
\newblock Item {Response} {Theory}: {What} {It} {Is} and {How} {You} {Can}
  {Use} the {IRT} {Procedure} to {Apply} {It}.

\bibitem[Bandura, 2001]{bandura_social_2001}
Bandura, A. (2001).
\newblock Social cognitive theory: {An} agentic perspective.
\newblock {\em Annual Review of Psychology}, 52(1):1--26.

\bibitem[Bills, 2003]{bills_credentials_2003}
Bills, D.~B. (2003).
\newblock Credentials, {Signals}, and {Screens}: {Explaining} the
  {Relationship} {Between} {Schooling} and {Job} {Assignment}.
\newblock {\em Review of Educational Research}, 73(4):441--469.

\bibitem[Black and Wiliam, 1998]{black_assessment_1998}
Black, P. and Wiliam, D. (1998).
\newblock Assessment and {Classroom} {Learning}.
\newblock {\em Assessment in Education: Principles, Policy \& Practice},
  5(1):7--74.

\bibitem[Ciarli et~al., 2021]{ciarli_digital_2021}
Ciarli, T., Kenney, M., Massini, S., and Piscitello, L. (2021).
\newblock Digital technologies, innovation, and skills: {Emerging} trajectories
  and challenges.
\newblock {\em Research Policy}, 50(7):104289.

\bibitem[Corbett and Anderson, 1995]{corbett_knowledge_1995}
Corbett, A.~T. and Anderson, J.~R. (1995).
\newblock Knowledge tracing: {Modeling} the acquisition of procedural
  knowledge.
\newblock {\em User Modelling and User-Adapted Interaction}, 4(4):253--278.

\bibitem[Da~Silva et~al., 2023]{da_silva_systematic_2023}
Da~Silva, F.~L., Slodkowski, B.~K., Da~Silva, K. K.~A., and Cazella, S.~C.
  (2023).
\newblock A systematic literature review on educational recommender systems for
  teaching and learning: research trends, limitations and opportunities.
\newblock {\em Education and Information Technologies}, 28(3):3289--3328.

\bibitem[Deci and Ryan, 2000]{deci_what_2000}
Deci, E.~L. and Ryan, R.~M. (2000).
\newblock The "{What}" and "{Why}" of {Goal} {Pursuits}: {Human} {Needs} and
  the {Self}-{Determination} of {Behavior}.
\newblock {\em Psychological Inquiry}, 11(4):227--268.

\bibitem[Deschênes, 2020]{deschenes_recommender_2020}
Deschênes, M. (2020).
\newblock Recommender systems to support learners’ {Agency} in a {Learning}
  {Context}: a systematic review.
\newblock {\em International Journal of Educational Technology in Higher
  Education}, 17(1):50.

\bibitem[{European Centre for the Development of Vocational Training.},
  2025]{european_centre_for_the_development_of_vocational_training_cedefop_2025}
{European Centre for the Development of Vocational Training.} (2025).
\newblock {\em Cedefop labour and skills shortage index.}
\newblock Publications Office, LU.

\bibitem[Graesser et~al., 2004]{graesser_autotutor_2004}
Graesser, A.~C., Lu, S., Jackson, G.~T., Mitchell, H.~H., Ventura, M., Olney,
  A., and Louwerse, M.~M. (2004).
\newblock {AutoTutor}: {A} tutor with dialogue in natural language.
\newblock {\em Behavior Research Methods, Instruments, \& Computers},
  36(2):180--192.

\bibitem[Kizilcec et~al., 2017]{kizilcec_self-regulated_2017}
Kizilcec, R.~F., Pérez-Sanagustín, M., and Maldonado, J.~J. (2017).
\newblock Self-regulated learning strategies predict learner behavior and goal
  attainment in {Massive} {Open} {Online} {Courses}.
\newblock {\em Computers \& Education}, 104:18--33.

\bibitem[Memarian and Doleck, 2025]{memarian_education_2025}
Memarian, B. and Doleck, T. (2025).
\newblock Education with a systems theory and control perspective.
\newblock {\em Education and Information Technologies}, 30(9):12249--12266.

\bibitem[Pasquale et~al., 2026]{pasquale_recommender_2026}
Pasquale, G.~D., Dean, S., and Frasca, P. (2026).
\newblock Recommender {Systems} as {Control} {Systems}.
\newblock arXiv:2605.01503 [eess.SY].

\bibitem[Ricci et~al., 2022]{ricci_recommender_2022}
Ricci, F., Rokach, L., and Shapira, B., editors (2022).
\newblock {\em Recommender {Systems} {Handbook}}.
\newblock Springer, New York, NY, third edition 2022 edition.

\bibitem[Roediger and Karpicke, 2006]{roediger_test-enhanced_2006}
Roediger, H.~L. and Karpicke, J.~D. (2006).
\newblock Test-{Enhanced} {Learning}: {Taking} {Memory} {Tests} {Improves}
  {Long}-{Term} {Retention}.
\newblock {\em Psychological Science}, 17(3):249--255.

\bibitem[Schaffter and Bounekkar, 2026]{schaffter_agency_2026}
Schaffter, E. and Bounekkar, A. (2026).
\newblock Agency and {Legitimacy}: a quadrant of 30 digital educational
  systems.
\newblock Supplementary material, CC BY 4.0 (data) and MIT (code).

\bibitem[Schaffter et~al., 2025]{schaffter_trajectory-based_2025}
Schaffter, E., Bounekkar, A., and Negre, E. (2025).
\newblock Trajectory-{Based} {Recommender} {Systems} as {Control} {Systems}.
\newblock {\em ICAT 2025}.

\bibitem[VanLehn, 2006]{vanlehn_behavior_2006}
VanLehn, K. (2006).
\newblock The {Behavior} of {Tutoring} {Systems}.
\newblock {\em International Journal of Artificial Intelligence in Education},
  16(3):227--265.

\bibitem[VanLehn, 2011]{vanlehn_relative_2011}
VanLehn, K. (2011).
\newblock The {Relative} {Effectiveness} of {Human} {Tutoring}, {Intelligent}
  {Tutoring} {Systems}, and {Other} {Tutoring} {Systems}.
\newblock {\em Educational Psychologist}, 46(4):197--221.

\bibitem[Wang et~al., 2025]{wang_development_2025}
Wang, X., Maeda, Y., and Chang, H.-H. (2025).
\newblock Development and techniques in learner model in adaptive e-learning
  system: {A} systematic review.
\newblock {\em Computers \& Education}, 225:105184.

\bibitem[Weber, 1978]{weber_economy_1978}
Weber, M. (1978).
\newblock {\em Economy and {Society}: {An} {Outline} of {Interpretive}
  {Sociology}}.
\newblock University of California Press, Berkeley.
\newblock Original work published 1922 as \textit{Wirtschaft und Gesellschaft}.

\bibitem[Wrosch et~al., 2003]{wrosch_adaptive_2003}
Wrosch, C., Scheier, M.~F., Miller, G.~E., Schulz, R., and Carver, C.~S.
  (2003).
\newblock Adaptive {Self}-{Regulation} of {Unattainable} {Goals}: {Goal}
  {Disengagement}, {Goal} {Reengagement}, and {Subjective} {Well}-{Being}.
\newblock {\em Personality and Social Psychology Bulletin}, 29(12):1494--1508.

\bibitem[Zimmerman, 2002]{zimmerman_becoming_2002}
Zimmerman, B.~J. (2002).
\newblock Becoming a {Self}-{Regulated} {Learner}: {An} {Overview}.
\newblock {\em Theory Into Practice}, 41(2):64--70.

\end{thebibliography}
}

\end{document}